%Paper: hep-th/9506144
%From: ANNA CERESOLE 3911564 7358 <CERESOLE@polito.it>
%Date: Wed, 21 Jun 1995 17:57:37 GMT+1
%Date (revised): Wed, 21 Jun 1995 22:05:18 GMT+1

%%%%%%%%%%%%%%%%%%%%%%%%%%%%%%%%%%%%%%%%%%%%%%%%%%%%%%%%%%%%
\input harvmac
%%%%%%%%%%%%%%%%%%%%%%%%%%%%%%%%%%%%%%%%%%%%%%%%%%%%%%%%%%%%%
\newif\ifdraft

\noblackbox
\catcode`\@=11
\newif\iffrontpage
%%%%%%%%%%%%%%%%%%%%%%%%%%%%%%%%%%%%%%%%%%%%%%%%%%%%%%%%%%%%%%
%%%%% sizes, offsets etc
%%%%%%%%%%%%%%%%%%%%%%%%%%%%%%%%%%%%%%%%%%%%%%%%%%%%%%%%%%%%%%
\ifx\answ\bigans
\def\titleft{\titsm}
\magnification=1200\baselineskip=14pt plus 2pt minus 1pt
%
%%%%% unreduced mode: %%%%
%\voffset=0.35truein\hoffset=0.250truein
\advance\hoffset by-0.075truein
\hsize=6.15truein\vsize=600.truept\hsbody=\hsize\hstitle=\hsize
\else\let\lr=L
\def\titleft{\titla}
\magnification=1000\baselineskip=14pt plus 2pt minus 1pt
%
%%%%% reduced mode: %%%%%%%
%\hoffset=-0.5truein\voffset=-.1truein
\hoffset=-.48truein\voffset=-.1truein
\vsize=6.5truein
\hstitle=8.truein\hsbody=4.75truein
\fullhsize=10truein\hsize=\hsbody
\fi
%
%\hstitle=8.truein\hsbody=4.5truein
%\fullhsize=9.75truein\hsize=\hsbody
\parskip=4pt plus 15pt minus 1pt
%%%%%%%%%%%%%%%%%%%%%%%%%%%%%%%%%%%%%%%%%%%%%%%%%%%%%%%%%%%%%%
%%%%%  fonts
%%%%%%%%%%%%%%%%%%%%%%%%%%%%%%%%%%%%%%%%%%%%%%%%%%%%%%%%%%%%%%
%%%%%%%%%%%%%%%%%%%%%%%%%%%%%%%%%%%%%%%%%%%%%%%%%%%%%%%%%%%%%%

\font\titla=cmr10 scaled\magstep3
\font\tenmss=cmss10
\font\absmss=cmss10 scaled\magstep1

\font\twelvebf=cmbx10 scaled\magstep1

\newfam\mssfam
\font\footrm=cmr8  \font\footrms=cmr5
\font\footrmss=cmr5   \font\footi=cmmi8
\font\footis=cmmi5   \font\footiss=cmmi5
\font\footsy=cmsy8   \font\footsys=cmsy5
\font\footsyss=cmsy5   \font\footbf=cmbx8
\font\footmss=cmss8
\def\footfont{\def\rm{\fam0\footrm}
\textfont0=\footrm \scriptfont0=\footrms
\scriptscriptfont0=\footrmss
\textfont1=\footi \scriptfont1=\footis
\scriptscriptfont1=\footiss
\textfont2=\footsy \scriptfont2=\footsys
\scriptscriptfont2=\footsyss
\textfont\itfam=\footi \def\it{\fam\itfam\footi}
\textfont\mssfam=\footmss \def\mss{\fam\mssfam\footmss}
\textfont\bffam=\footbf \def\bf{\fam\bffam\footbf} \rm}
\def\tenpoint{\def\rm{\fam0\tenrm}
\textfont0=\tenrm \scriptfont0=\sevenrm
\scriptscriptfont0=\fiverm
\textfont1=\teni  \scriptfont1=\seveni
\scriptscriptfont1=\fivei
\textfont2=\tensy \scriptfont2=\sevensy
\scriptscriptfont2=\fivesy
\textfont\itfam=\tenit \def\it{\fam\itfam\tenit}
\textfont\mssfam=\tenmss \def\mss{\fam\mssfam\tenmss}
\textfont\bffam=\tenbf \def\bf{\fam\bffam\tenbf} \rm}
\ifx\answ\bigans\def\abstractfont{\tenpoint}\else
\def\abstractfont{\def\rm{\fam0\absrm}
\textfont0=\absrm \scriptfont0=\absrms
\scriptscriptfont0=\absrmss
\textfont1=\absi \scriptfont1=\absis
\scriptscriptfont1=\absiss
\textfont2=\abssy \scriptfont2=\abssys
\scriptscriptfont2=\abssyss
\textfont\itfam=\bigit \def\it{\fam\itfam\bigit}
\textfont\mssfam=\absmss \def\mss{\fam\mssfam\absmss}
\textfont\bffam=\absbf \def\bf{\fam\bffam\absbf}\rm}\fi
%
%%%%%%%%%%%%%%%%%%%%%%%%%%%%%%%%%%%%%%%%%%%%%%%%%%%%%%%%%%%%%%
%%%%% footnotes   (adapted from PHYZZX)
%%%%%%%%%%%%%%%%%%%%%%%%%%%%%%%%%%%%%%%%%%%%%%%%%%%%%%%%%%%%%%
\def\f@@t{\baselineskip10pt\lineskip0pt\lineskiplimit0pt
\bgroup\aftergroup\@foot\let\next}
\setbox\strutbox=\hbox{\vrule height 8.pt depth 3.5pt width\z@}
\def\vfootnote#1{\insert\footins\bgroup
\baselineskip10pt\footfont
\interlinepenalty=\interfootnotelinepenalty
\floatingpenalty=20000
\splittopskip=\ht\strutbox \boxmaxdepth=\dp\strutbox
\leftskip=24pt \rightskip=\z@skip
\parindent=12pt \parfillskip=0pt plus 1fil
\spaceskip=\z@skip \xspaceskip=\z@skip
\Textindent{$#1$}\footstrut\futurelet\next\fo@t}
\def\Textindent#1{\noindent\llap{#1\enspace}\ignorespaces}
\def\footnote#1{\attach{#1}\vfootnote{#1}}%

\def\foot{\attach\footsymbolgen\vfootnote{\footsymbol}}
\let\footsymbol=\star
\newcount\lastf@@t           \lastf@@t=-1
\newcount\footsymbolcount    \footsymbolcount=0
\def\footsymbolgen{\relax\footsym
\global\lastf@@t=\pageno\footsymbol}
\def\footsym{\ifnum\footsymbolcount<0
\global\footsymbolcount=0\fi
{\iffrontpage \else \advance\lastf@@t by 1 \fi
\ifnum\lastf@@t<\pageno \global\footsymbolcount=0
\else \global\advance\footsymbolcount by 1 \fi }
\ifcase\footsymbolcount \fd@f\star\or
\fd@f\dagger\or \fd@f\ast\or
\fd@f\ddagger\or \fd@f\natural\or
\fd@f\diamond\or \fd@f\bullet\or
\fd@f\nabla\else \fd@f\dagger
\global\footsymbolcount=0 \fi }
\def\fd@f#1{\xdef\footsymbol{#1}}
\def\space@ver#1{\let\@sf=\empty \ifmmode #1\else \ifhmode
\edef\@sf{\spacefactor=\the\spacefactor}
\unskip${}#1$\relax\fi\fi}
\def\attach#1{\space@ver{\strut^{\mkern 2mu #1}}\@sf}
%
%%%%%%%%%%%%%%%%%%%%%%%%%%%%%%%%%%%%%%%%%%%%%%%%%%%%%%%%%%%%%%
%%%%% References
%%%%%%%%%%%%%%%%%%%%%%%%%%%%%%%%%%%%%%%%%%%%%%%%%%%%%%%%%%%%%%
\newif\ifnref
\def\rrr#1#2{\relax\ifnref\nref#1{#2}\else\ref#1{#2}\fi}
\def\ldf#1#2{\begingroup\obeylines
\gdef#1{\rrr{#1}{#2}}\endgroup\unskip}
\def\nrf#1{\nreftrue{#1}\nreffalse}
\def\doubref#1#2{\refs{{#1},{#2}}}

\nreffalse
\def\refout{\listrefs}
%
%%%%%%%%%%%%%%%%%%%%%%%%%%%%%%%%%%%%%%%%%%%%%%%%%%%%%%%%%%%%%%
%%%%%%% eq numbering
%%%%%%%%%%%%%%%%%%%%%%%%%%%%%%%%%%%%%%%%%%%%%%%%%%%%%%%%%%%%%%
\def\eqn#1{\xdef #1{(\secsym\the\meqno)}
\writedef{#1\leftbracket#1}%
\global\advance\meqno by1\eqno#1\eqlabeL#1}
\def\eqnalign#1{\xdef #1{(\secsym\the\meqno)}
\writedef{#1\leftbracket#1}%
\global\advance\meqno by1#1\eqlabeL{#1}}
%
%%%%%%%%%%%%%%%%%%%%%%%%%%%%%%%%%%%%%%%%%%%%%%%%%%%%%%%%%%%%%%
%%%%%%  macros for titlepage, marginnotes, etc
%%%%%%%%%%%%%%%%%%%%%%%%%%%%%%%%%%%%%%%%%%%%%%%%%%%%%%%%%%%%%%
\def\chap#1{\global\advance\secno by1\message{(\the\secno\ #1)}
%\ifx\answ\bigans \vfill\eject \else \bigbreak\bigskip \fi  %if
% desired
\global\subsecno=0\eqnres@t\noindent{\twelvebf\the\secno\ #1}
\writetoca{{\secsym} {#1}}\par\nobreak\medskip\nobreak}
%% FOLLOWING LINE CANNOT BE BROKEN BEFORE 70 CHAR
%% FOLLOWING LINE CANNOT BE BROKEN BEFORE 70 CHAR
%% FOLLOWING LINE CANNOT BE BROKEN BEFORE 70 CHAR
\def\eqnres@t{\xdef\secsym{\the\secno.}\global\meqno=1\bigbreak\bigskip}
\def\sequentialequations{\def\eqnres@t{\bigbreak}}\xdef\secsym{}
\global\newcount\subsecno \global\subsecno=0
\def\sect#1{\global\advance\subsecno
by1\message{(\secsym\the\subsecno. #1)}
\ifnum\lastpenalty>9000\else\bigbreak\fi
\noindent{\bf\secsym\the\subsecno\ #1}\writetoca{\string\quad
{\secsym\the\subsecno.} {#1}}\par\nobreak\medskip\nobreak}
%%%%%%%%%%%%%%%%%%%%%%%%%%%%%%%%%%%%%%%%%%%%%%%%%%%%%%%%%%%%%%
%\def\chap#1{\newsec{#1}}
\def\chapter#1{\chap{#1}}
\def\section#1{\sect{#1}}
\def\\{\ifnum\lastpenalty=-10000\relax
\else\hfil\penalty-10000\fi\ignorespaces}
\def\note#1{\leavevmode%
\edef\@@marginsf{\spacefactor=\the\spacefactor\relax}%
\ifdraft\strut\vadjust{%
\hbox to0pt{\hskip\hsize%
\ifx\answ\bigans\hskip.1in\else\hskip .1in\fi%
\vbox to0pt{\vskip-\dp
%\vskip4pt
\strutbox\sevenbf\baselineskip=8pt plus 1pt minus 1pt%
\ifx\answ\bigans\hsize=.7in\else\hsize=.35in\fi%
\tolerance=5000 \hbadness=5000%
\leftskip=0pt \rightskip=0pt \everypar={}%
\raggedright\parskip=0pt \parindent=0pt%
\vskip-\ht\strutbox\noindent\strut#1\par%
\vss}\hss}}\fi\@@marginsf\kern-.01cm}
\def\titlepage{%
\frontpagetrue\nopagenumbers\abstractfont%
\hsize=\hstitle\rightline{\vbox{\baselineskip=10pt%
{\abstractfont\pubnum}}}\pageno=0}
\frontpagefalse
\def\pubnum{}
\def\pdate{\number\month/\number\yearltd}
\def\makefootline{\iffrontpage\vskip .27truein
\line{\the\footline}
%\vskip -.1truein\line{\pdate\hfil}
\vskip -.1truein\leftline{\vbox{\baselineskip=10pt%
{\abstractfont\pdate}}}
\else\vskip.5cm\line{\hss \tenrm $-$ \folio\ $-$ \hss}\fi}
\def\title#1{\vskip .7truecm\titlestyle{\titleft #1}}
\def\titlestyle#1{\par\begingroup \interlinepenalty=9999
\leftskip=0.02\hsize plus 0.23\hsize minus 0.02\hsize
\rightskip=\leftskip \parfillskip=0pt
\hyphenpenalty=9000 \exhyphenpenalty=9000
\tolerance=9999 \pretolerance=9000
\spaceskip=0.333em \xspaceskip=0.5em
\noindent #1\par\endgroup }
\def\autskip{\ifx\answ\bigans\vskip.5truecm\else\vskip.1cm\fi}
\def\author#1{\vskip .7in \centerline{#1}}

\def\address#1{\ifx\answ\bigans\vskip.2truecm
\else\vskip.1cm\fi{\it \centerline{#1}}}
\def\abstract#1{
\vskip .5in\vfil\centerline
{\bf Abstract}\penalty1000
{{\smallskip\ifx\answ\bigans\leftskip 2pc \rightskip 2pc
\else\leftskip 5pc \rightskip 5pc\fi
\noindent\abstractfont \baselineskip=12pt
{#1} \smallskip}}
\penalty-1000}
\def\endpage{\tenpoint\supereject\global\hsize=\hsbody%
\frontpagefalse\footline={\hss\tenrm\folio\hss}}
\def\ack{\goodbreak\vskip2.cm\centerline{{\bf Acknowledgements}}}
%%%%%%%%%%%%%%%%%%%%%%%%%%%%%%%%%%%%%%%%%%%%%%%%%%%%%%%%%%%%%%
\def\a{\alpha} \def\b{\beta} \def\d{\delta}

\def\L{\Lambda} 
\def\cA{{\cal A}} 
 
\def\cF{{\cal F}} 
 
 \def\cK{{\cal K}}
\def\cL{{\cal L}}

 \def\cV{{\cal V}}

\def\IGa{\ralax{{\rm I}\kern-.18em \Gamma}}
\def\IZ{{\hbox{{\rm Z}\kern-.4em\hbox{\rm Z}}}}
\def\IR{{\hbox{{\rm I}\kern-.4em\hbox{\rm R}}}}
\def\nup#1({Nucl.\ Phys.\ $\us {B#1}$\ (}
\def\plt#1({Phys.\ Lett.\ $\us  {#1}$\ (}
\def\cmp#1({Comm.\ Math.\ Phys.\ $\us  {#1}$\ (}
\def\prp#1({Phys.\ Rep.\ $\us  {#1}$\ (}
\def\prl#1({Phys.\ Rev.\ Lett.\ $\us  {#1}$\ (}
\def\prv#1({Phys.\ Rev.\ $\us  {#1}$\ (}
\def\mpl#1({Mod.\ Phys.\ Let.\ $\us  {A#1}$\ (}
\def\ijmp#1({Int.\ J.\ Mod.\ Phys.\ $\us {A#1}$\ (}
\def\cqg#1({Class.\ Quantum Grav.\ $\us {#1}$\ (}
\def\anp#1({Ann.\ of Phys.\ $\us {#1}$\ (}
\def\tmp#1({Theor.\ Math.\ Phys.\ $\us {#1}$\ (}
\def\tit#1|{{\it #1},\ }
%
%%%%%%%%%%%%%%%%%%%%%%%%%%%%%%%%%%%%%%%%%%%%%%%%%%%%%%%%%%%%%%
%%%%% misc macros %%%%%
%%%%%%%%%%%%%%%%%%%%%%%%%%%%%%%%%%%%%%%%%%%%%%%%%%%%%%%%%%%%%%
\def\ni{\noindent}

\def\bar{\overline}
\def\us#1{\underline{#1}}

\def\hat{\widehat}

\def\notin{\hbox{{$\in$}\kern-.51em\hbox{/}}}

\def\del{\partial}

\def\eg{{\it e.g.}\ } \def\ie{{\it i.e.}\ }
\catcode`\@=12

\def\K{K\"ahler\ }

%%%%%%%%%%%%%%%%%FRONTPAGE%%%%%%%%%%%%%%%%%%%%%%%%%%%%%%%%%%%%%%%%%%%

\def\aff#1#2{\centerline{$^{#1}${\it #2}}}
\def\pubnum{
\hbox{CERN-TH/95-166}
\hbox{POLFIS-TH. 08/95}
\hbox{UCLA/95/TEP/23}
\hbox{hep-th/9506144}}
\titlepage
\vskip .5truecm
\title
 {11-Dimensional Supergravity Compactified on Calabi--Yau Threefolds
\foot{Supported in part by DOE
 grant DE-FGO3-91ER40662,Task C. and by EEC
 Science Program SC1*CT92-0789.}}
\vskip-.2cm
\author{A.\ C.\ Cadavid$^{1}$,
A.\ Ceresole$^{2}$, R.\ D'Auria$^{2}$ and
S.\ Ferrara$^{3}$}
\vskip2.truecm
\aff1{Department of Physics and Astronomy, California State University,}
\centerline{Northridge, CA 91330, USA}
\line{\hfill}
\aff2{Dipartimento di Fisica, Politecnico di Torino,}
\centerline{\it  Corso Duca Degli Abruzzi 24, 10129 Torino, Italy}
\centerline{\it and}
\centerline{\it INFN, Sezione di Torino, Italy}
\line{\hfill}
\aff3{CERN, 1211 Geneva 23, Switzerland}
%\centerline{\it and}
%\centerline{\it UCLA Physics Department, Los Angeles CA 90024-1547, USA}
\line{\hfill}
%%%%%%%%%%%%%%%%%%%%%%%%%%%%%%%%%%%%%%%%%%%%%%%%%%%%%%%%%%%%%%%%%
\vskip-.8 cm
\def\abs
{\ni
We consider generic features of eleven dimensional supergravity
compactified down to five dimensions on an arbitrary Calabi--Yau threefold.
}
\abstract{\abs}
\vfill
\endpage
\baselineskip=14pt plus 2pt minus 1pt
%%%%%%%%%%%%%%%%%%%%%%%%%%%%%%%%%%%%%%%%%%%%%%%%%%%%%%%%%%%%%%%%%%%%%%%%
%%%%%%%%%%%%REFERENCES%%%%%%%%%%%%%%%%%%%%%%%%%%%%%%%%%%%%%%%%%%%%%%%%%%
\ldf\huto{C. Hull and P. K. Townsend, preprint QMW-94-30, R/94/33.
hep-th/9501030.}
\ldf\duffb{M.\ Duff, preprint NI-94-033, CTP-TAMU-49/94,
hep-th/9501030.}
\ldf\town{P. K. Townsend, preprint hep-th/9501068, R/52/2.}
\ldf\SW{
N.~Seiberg and E.~Witten  , \nup426 (1994) 19; \nup431 (1994) 484.}
\ldf\BSV{E. Bergshoeff, E. Sezgin and A. Van Proeyen, \nup264 (1986) 653.}
\ldf\cande{P. Candelas and X. de la Ossa, \nup355 (1991) 455.}
\ldf\kltold{ A.~Klemm, W.~Lerche, S.~Theisen and
S.~Yankielovicz, \plt344 (1995) 169.}
% \tit On the Monodromies of
%N=2 Supersymmetric Yang--Mills Theory|
%Proceedings of the Workshop on \tit
%Physics from the Planck
%Scale to Electromagnetic Scale| Warsaw 1994 and
%of the \tit 28th International
%Symposium on Particle Theory|  Wendisch-Rietz,
%preprint CERN-TH.7538/94,
%hep-th/9412158.}

\ldf\faraggi{ P. Argyres and A. Faraggi, \prl73 (1995) 3931.}

\ldf\CDF{
A. Ceresole, R. D'Auria and S. Ferrara,\plt339B (1994) 71.}

\ldf\AFGNT{
I.~Antoniadis, S.~Ferrara, E.~Gava, K.\ S.\  Narain and T.\ R.\
  Taylor, \tit Perturbative Prepotential and Monodromies in N=2
Heterotic Strings| preprint hep-th/9504034.}

\ldf\DWKLL{B.\ de Wit, V.\ Kaplunovsky, J.\ Louis and D.\ L\"ust,
\tit Perturbative Couplings of Vector Multiplets in N=2 Heterotic String
 Vacua| preprint hep-th/9504006.}

\ldf\HS{
J.\ A.\ Harvey and A.\ Strominger,
 ``The Heterotic String is a Soliton'', preprint hep-th/9504047.}

\ldf\stromco{
A.~Strominger,
``Massless Black Holes and Conifolds in String Theories", hep-th/9504090;
B.~Greene, D.~Morrison and A.~Strominger,
 \tit Black hole condensation and the unification of string
vacua| preprint hep-th/9504145.}

\ldf\lopez{ C. Gomez and E. Lopez,\tit From Quantum Monodromy
to Duality| preprint hep-th 9505135.}
\ldf\lopezb{ C. Gomez and E. Lopez,
\tit A Note on the String Analog of N=2 Supersymmetric Yang-Mills|
 preprint hep-th/9506024.}

\ldf\kltnew{
A.~Klemm, W.~Lerche and S.~Theisen, \tit Nonperturbative Effective
Actions of N=2 Supersymmetric Gauge Theories| preprint CERN-TH/95-104,
LMU-TPW 95-7, hep-th/9505150.}

\ldf\KLM{A. Klemm, W. Lerche and P. Mayr,
\tit K3-Fibrations and Heterotic-Type II Strings Duality|
 preprint CERN-TH/95-165, hep-th/9506112.}
%%%%%%%%%%%%%%%%%%%%%%%%%%%%%%%%%%%%%%%%%%%%%%%%%%%%%%%%%%%%%%%
%
\ldf\csf{E.\ Cremmer, J.\ Scherk and S.\ Ferrara, Phys. Lett.
${\us {74B}}$ (1978) 61.}
\ldf\susu{B.\ de Wit and A.\ van Proeyen, \nup245 (1984) 89;
E.\ Cremmer, C.\ Kounnas, A.\ van Proeyen, J.\ Derendinger, S.\ Ferrara,
B.\ De Wit and L.\ Girardello, \nup250 (1985) 385;
B.\ de Wit, P.\ G.\ Lauers and A.\ Van Proeyen, \nup255 (1985) 569;
S.\ Cecotti, S.\ Ferrara and L.\ Girardello, \ijmp4 (1989) 2475.}
%%%%%%%%%%%%%%%%%%%%%%%%%%%%%%%%%%%%%%%%%%%%%%%%%%%%%%%%%%%%%%%%%%%%%%
\ldf\shsh{ J. Sherk and J. H. Schwarz, \plt82B (1979)70;\nup153 (198) 61.}
\ldf\gmass{E. Cremmer, J. Scherk and J. H.  Schwarz,
Phys. Lett. ${\us {84B}}$ (1979) 83; S. Ferrara and B.  Zumino,
Phys. Lett. ${\us {86B}}$ (1979) 279.}

\ldf\hawk{see \eg E. Cremmer, in \tit Superspace and Supergravity|
edited by S.E Hawking and M. Rocek,  Cambridge University Press
 (1981) 267.}

\ldf\rgm{R. D'Auria, P. Fr\`e , E. Maina and T. Regge, Ann. of Phys.
135 (1981) 237.}

\ldf\fkpz{S. Ferrara, C. Kounnas, M. Porrati and F. Zwirner,
\nup318 (1989) 75.}
\ldf\wit{E. Witten, \plt155B (1985) 151.}

\ldf\CDFVP{
A.~Ceresole, R.~D'Auria, S.~Ferrara and A.~Van Proeyen,
\tit Duality Transformations in Supersymmetric Yang--Mills
Theories coupled to Supergravity| preprint CERN-TH. 7547/94,
hep-th/9502072 (1995), Nucl. Phys.  B, to be published;
\tit On Electromagnetic Duality in Locally
Supersymmetric N=2 Yang--Mills Theory| preprint CERN-TH.08/94,
hep-th/9412200, Proceedings of the Workshop on \tit  Physics from the Planck
Scale to Electromagnetic Scale| Warsaw 1994.}

\ldf\Wdy{
E.~Witten, \tit String Theory Dynamics in Various Dimensions|
IASSNS-HEP-95-18, hep-th/9503124.}

\ldf\FHSV{
S.\ Ferrara, J.\ A.\ Harvey, A.\ Strominger and C.\ Vafa,
\tit Second-quantized Mirror Symmetry| preprint  hep-th/9505162.}

\ldf\kava{
S.~Kachru and C.~Vafa, \tit Exact Results for N=2
Compactifications of Heterotic Strings| preprint
 HUTP-95/A016, hep-th/9505105.
}

\ldf\noi{ M. Bill\'o, A. Ceresole, R. D'Auria, S. Ferrara, P. Fr\`e,  P.
Soriani and A. Van Proeyen, \tit A Search for Nonperturbative Dualities of
Local N=2 Yang--Mills Theories from Calabi--Yau Threefolds| preprint
hep-th/9506075.}

\ldf\CFG{ S.\ Cecotti, S.\ Ferrara, L.\ Girardello, \ijmp4
(1989) 2475; \plt213 (1988) 443.}

\ldf\mirror{ see \eg \tit Essays on Mirror Manifolds| ed. S.T. Yau
 (Int. Press, Honk Kong, 1992).}

\ldf\BCF{ M. Bodner, A. C. Cadavid and S. Ferrara,
\cqg8 (1991) 789.}

\ldf\salsez{See \eg  \tit Supergravities in Diverse Dimensions|
A. Salam and E. Sezgin eds., World Scientific (1989).}

\ldf\duff{Compactification of D=11 Supergravity on a particular Calabi--Yau
threefold was first discussed by M. Duff,
\tit Architecture of Fundamental Interactions at Short Distances|
Les Houches Lectures (1985) 819,
 P. Ramond and R. Stora edts., North Holland.}

\ldf\gst{M. G\"unaydin, G. Sierra and P. K. Townsend, \nup242 (1984) 244;
\nup253 (1985) 573.}

\ldf\strom{A. Strominger, \prl55 (1985) 2547.}

\ldf\cjs{E. Cremmer, B. Julia and J. Scherk, Phys. Lett.
${\us {76B}}$ (1978) 409.}

\ldf\fesaba{S. Ferrara and S. Sabharwal, \cqg6 (1989) 277.}

\ldf\fesabb{S. Ferrara and S. Sabharwal, \nup332 (1990) 317.}

\ldf\cremet{E. Cremmer, C. Kounnas, A. van Proeyen, J. P. Derendinger,
S. Ferrara, B. De Wit and L. Girardello, \nup250 (1985) 385.}

\ldf\roma{L. Romans, \nup276 (1986) 71.}

\ldf\seib{N. Seiberg, \nup303 (1988) 286.}

\ldf\nara{K. Narain, \plt169B (1986) 41; K. Narain, M. Sarmadi
and E. Witten, \nup279 (1987) 369.}

%%%%%%%%%%%%%%%%%%%%%%%%%%%%%%%%%%%%%%%%%%%%%%%%%%%%%%%%%%%%%%%%%%%%%%%%
%%%%%%%%%%%%%%%%%%%%%%%%%%%%%%%BODY%%%%%%%%%%%%%%%%%%%%%%%%%%%%%%%%%%%%%
\def\L{{\Lambda}}
\def\S{{\Sigma}}
\def\D{{\Delta}}
\def\cA{{\cal A}}
\def\cF{{\cal F}}
\def\cV{{\cal V}}
\def\bi{{\bar\imath}}
\def\bj{{\bar\jmath}}

Recently, it has been suggested that eleven dimensional supergravity
may arise as an effective theory of some string theories in their strong
coupling regime \Wdy . In particular, compactification of $D=11$
supergravity to diverse dimensions $D<11$ contains a Kaluza--Klein spectrum
which is a natural candidate, as shown by Witten, for some non perturbative
BPS states of string theories. In this note we fill a gap in this analysis
by considering $D=11$ supergravity compactified to five dimensions on an
arbitrary Calabi--Yau manifold with Hodge numbers $(h_{(1,1)},h_{(2,1)})$ and
intersection matrix $d_{\L\S\D}$ ($\L,\S,\D=1,\ldots, h_{(1,1)}$). Here we will
only report the generic structure which emerges in doing this analysis,
while the complete action of the theory will be given elsewhere.

The five dimensional theory obtained in this way  happens
to contain the gravity multiplet\salsez\
$$
(e_{a\mu},\psi_{\mu I}, \cA_\mu)\ \ \ \ \ \ \ (I=1,2)\ ,
\eqn\uno
$$
{}~$h_{(1,1)}-1$ vector multiplets
$$
(\cA_\mu^A,\lambda^A_I,\phi^A)\ \ \ \ \  \ (A=1,\ldots,h_{(1,1)}-1)\ ,
\eqn\due
$$
and $h_{(2,1)}+1$ hypermultiplets
$$
(\zeta^m,\cA^m_I)\ \ \ \ \ \ (m=1,\ldots,2(h_{(2,1)}+1))\ .
\eqn\tre
$$
It is convenient to introduce a vector index $\L=1,\ldots,h_{(1,1)}$
 which covers
also the graviphoton. Then, the entire coupling of vector multiplets
to five dimensional supergravity is specified, as shown in \gst ,
by the intersection numbers $d_{\L\S\D}$ which, in particular, express
 the coupling of the 5D topological term
$$
\int d^5x\ d_{\L\S\D} \cF^\L\wedge \cF^\S \wedge \cA^\D \ .
\eqn\qua
$$

Before deriving the results, let us first show how the counting of degrees
of freedom for the bosonic fields is obtained. In $D=11$ we have a pure
geometrical theory (with no coupling constant) containing the metric
$G_{\hat\mu\hat\nu}$ and a three-form gauge field
$\cA_{\hat\mu\hat\nu\hat\rho}$. On a Calabi--Yau threefold with Hodge
numbers $(h_{(1,1)} , h_{(2,1)})$
 we obtain the following degrees of freedom\duff\
(the fermions, that we neglect here, just complete the multiplets),
splitting $\hat\mu=(\mu,i,\bi )$, ($\mu=1,\ldots,5, i,\bi =1,2,3$):
the graviton ($G_{\mu\nu}$),
$h_{(2,1)}$ complex scalars ($G_{ij}$), $h_{(1,1)}$ real scalars
($G_{i\bj}$), one real scalar ($\cA_{\mu\nu\rho}$),
$h_{(1,1)}$ vectors ($\cA_{\mu i \bj}$) , $h_{(2,1)}$
complex scalars ($\cA_{ij\bar k}$) and one complex scalar
 ($\cA_{ijk}=\epsilon_{ijk} C$).

\ni
So we get, as promised, $h_{(2,1)}+1$ hypermultiplet scalars
 $(G_{ij},\cA_{ij\bar k},\cA_{\mu\nu\rho},\cV, C)$,
$h_{(1,1)}-1$ vector multiplet scalars ($G_{i\bj}$ except the volume)
and $h_{(1,1)}$ vector fields $\cA_{\mu i \bj}$.

In the decomposition of the \K form\strom\
$$
J=\sum_{\L=1}^{h_{(1,1)}} M_\L\ V^\L \ \ \ \ \ (V^\L\in H^{(1,1)})
\eqn\cin
$$
one can extract the volume modulus
$$
\cV= {1\over{3!}}\int J \wedge J\wedge J
\eqn\sei
$$
and then consider moduli coordinates
$$
(t_\L ={{M_\L}\over{{\cV}^{1\over3}}},\cV)
\eqn\sette
$$
such that $\cV(t_\L)=1$. This is the natural splitting in five dimensions.
In fact, it is easy to see that $(\cV,\cA_{\mu\nu\rho},C)$ then becomes
an universal hypermultiplet, present in any Calabi--Yau compactification
\CFG , which of course has its counterpart in the dimensionally reduced theory
in $D=4$. If $h_{(2,1)}=0$, this multiplet belongs to the
${{SU(2,1)}\over{SU(2)\times U(1)}}$ quaternionic manifold, as shown in
\CFG .

In five dimensions, the vector multiplet moduli space will just be the
hypersurface $\cV=1$ of the classical \K cone\CFG ,
which is related to the moduli space of
the $M_\L$. This is precisely a space of the general form
allowed by 5D supergravity studied in \gst .
Note that the quantum moduli space for the $H^2$-cohomology (obtained by
mirror symmetry\mirror\ ) is not allowed by 5D supergravity\foot{The absence
of world-sheet instanton corrections to the \K metric in $D=5$ seems to be
related, by duality, to the absence of logarithmic infrared
singularities in the heterotic counterpart. Infact, an holomorphicity argument,
by further reduction to $D=4$, confirms the above statement.
This would give even more
evidence that world-sheet instantons on Calabi--Yau
\nrf{
\CDF\CDFVP\noi\kltold\stromco\kava\lopezb\KLM}\refs{\CDF{--}\KLM}
are dual, in $D=4$, to non perturbative singularities of $N=2$
microscopic Yang--Mills theories\SW (described by heterotic strings on
$N=2$ vacua).}
because of the
absence of the antisymmetric tensor $B_{\mu\nu}$. This is similar to the
analysis made in \Wdy , in the compactification of $D=11$ supergravity
on $K3$ down to $D=7$. There, the $K3$ moduli space was also the classical
one, \ie ${{SO(3,19)}\over{SO(3)\times SO(19)}}$ for precisely the same
 reason. On the other hand, the quaternionic manifold is compatible with
the moduli space of the complex structure of the Calabi--Yau and then will
in fact agree with what is usually obtained by the c-map
\doubref\CFG\fesabb .
Therefore, even in 5D, the quaternionic metric will be parametrized in terms
of the prepotential $F$ of the special geometry of the deformation of
the three-form complex cohomology. It is then obvious that the asymmetry
between $H^2$ and $H^3$ is a pure five dimensional phenomenon.

Let us briefly mention how the above results are actually derived. One
starts with the eleven dimensional geometrical theory of Cremmer, Julia
and Scherk\cjs
$$
\cL_{11} = -{1\over2} \hat e_{11} \hat R -{1\over{48}} \hat e
(\hat F_{\hat\mu_1\hat\mu_2\hat\mu_3\hat\mu_4})^2
+ {{\sqrt{2}}\over{12^4}}\epsilon^{\hat\mu_1\cdots\hat\mu_{11}}
\hat F_{\hat\mu_1\cdots\hat\mu_4}\hat F_{\hat\mu_5\cdots\hat\mu_8}
\hat A_{\hat\mu_9\cdots\hat\mu_{11}}\ .
\eqn\otto
$$
Compactification on a Calabi--Yau threefold is obtained, using the results
of ref. \BCF . For instance, from the Einstein term we obtain
$$\eqalign{
\cL_{5}= e_5 &\left\{-{1\over2}\cV_5 (M) R_5
+\cV_5 (M) \del_\mu z^\a \del_\mu \bar z^{\bar \b}\ G_{\a\bar\b}\right . \cr
 &\left . +{1\over2}\del_\mu M^\L \del_\mu M^\S
\left[ \cV_5 (M)G_{\L\S}+\cK_{\L\S}\right]\right\}\ ,\cr}
\eqn\nove
$$
where we have used the decomposition\cande\BCF
$$
\eqalign{
i\d G_{i\bar j} &= \sum_{\L=1}^{h_{(1,1)}}  M^\L V^\L_{i\bar j}\cr
{}~\d G_{ij} &= \sum_{\a=1}^{h_{(2,1)}}  \bar z^\a \bar b_{\a ij}\ ,\cr}
\eqn\dieci
$$
and we have defined
$$
\eqalign{
3! \cV(M) &\equiv \cK =d_{\L\S\D} M^\L M^\S M^\D\cr
G_{\L\S} &= -{1\over2}{\del\over{\del M^\L}}{\del\over{\del M^\S}} \log \cK=
-3 {{\cK_{\L\S}}\over{\cK}}+{9\over2}{{\cK_\L \cK_\S}\over{\cK^2}}\cr
\cK_\L &= d_{\L\S\D} M^\S M^\D \ \ \ ;\ \ \ \ \cK_{\L\S}=d_{\L\S\D}M^\D\cr
(G_{\L\S} M^\S &= {3\over2} {{\cK_\L}\over{\cK}}\ ,
\ G_{\L\S} M^\L M^\S ={3\over2}
)\ .\cr}
\eqn\undi
$$
By making a Weyl rescaling to bring the Einstein term to the canonical form,
the previous term becomes
$$\eqalign{
e_5^{-1} \cL_{5} &= -{{R_5}\over2}+G_{\a\bar\b}\del_\mu z^\a \del_\mu
\bar z^{\bar\b}-{1\over2} G_{\L\S}\del_\mu t^\L \del_\mu t^\S\cr
{}~&- {1\over4} (\del_\mu \log \cV_5(M))^2\cr}
\eqn\dodi
$$
with $\cV_5(t)=1$  and $M^\L=t^\L \cV_5(M)^{1/3}$.
The ($h_{(1,1)}-1$) $t^\L$ fields are precisely the coordinates
$\phi^A$ of the moduli space
of 5D supergravity with vector multiplet coupling fixed in terms of the
$d_{\L\S\D}$ symmetric symbols of ref. \gst . The other terms are part of
the hypermultiplets kinetic term, which now also include the volume modulus
$\cV(M)$.

It is easy to see, repeating a calculation similar to that of ref.
\fesaba\ , that in absence of the $z^\a$ scalars ($h_{(2,1)}=0$), the coupling
from the $\hat F$ four-forms, after dualizing $\cA_{\mu\nu\rho}$ to a
scalar field, reproduces, together with $\cA_{ijk}=\epsilon_{ijk} C$,
the one-dimensional quaternionic space ${{SU(2,1)}\over{SU(2)\times SU(1)}}$,
which was inferred in ref. \CFG\ and explicitly constructed in
refs. \doubref\fesaba\fesabb .

When the complex structure scalars are turned on, then one obtains a
quaternionic manifold identical to that discussed in refs.
\doubref\CFG\BCF  \foot{
We observe that the full lagrangian, including hypermultiplets,
 can be obtained in a straightforward
way either by reduction from $D=6$\BSV ,
or directly in $D=5$ using the techniques
developed in ref. \rgm . Quaternionic spaces encompassing
the dynamics of hypermultiplets are the same for $D=6,5$ and $4$.
However, one difference in the coupling to fermions is that in $D=5$,
being the scalar manifold real and the fermions non-chiral,
there is no \K connection in the covariant derivative of
fermions \hawk .}.

To make contact with string theories, we must proceed to further compactify
the 5D theory on $S_1$, so that we obtain a 4D theory which is the
compactification of $D=11$ supergravity on $CY\times S_1$\foot{ Notice that
in $D=4$ the $h_{(1,1)}$ moduli can be complexified by the additional $\cA_{5
i\bj}$ scalars, and a new vector multiplet comes from the metric massless
degrees of freedom $G_{5\mu},G_{55}$.}.

By introducing the $5$-dimensional radius $\phi_5$, and confining our
discussion to the $H^2$ moduli, we obtain after Weyl rescaling
$$
e^{-1} \cL_{4}=-{1\over2} R_4-{1\over2}
G_{\L\S} (t) \del_\mu t^\L \del_\mu t^\S
-{1\over{12}} \left[\del_\mu (\log \phi_5^3)\right]^2-{1\over4} (\del_\mu
\log \cV_5 (M))^2\ .
\eqn\tredi
$$
We can now compare this lagrangian with the one obtained by
compactifying 10D Type IIA theory on a Calabi--Yau threefold to $D=4$\BCF.
By splitting in this case the $H^2$ moduli in $(t^\L,\cV(v))$,
($\cV(t)=1$) we
get
$$
-{1\over2} R_4-{1\over2} G_{\L\S}(t) \del_\mu t^\L \del_\mu t^\S
-{1\over{12}} (\del_\mu \log \cV(v))^2
-{1\over4}\left[\del_\mu (\log \cV(v) \phi^{-3})\right]^2\ ,
\eqn\tordi
$$
where $\phi$ is the 10D dilaton field, related to the $10$ dimensional
Yang--Mills heterotic gauge coupling through the relation $g_{YM}^{-2}\sim
\phi^{-3/4}$.
We see that $\phi_5^3=\cV(v)$, while the
four dimensional dilaton is $\cV_5(M)$. By observing that\BCF
$$
v^\L=\phi^{3/4} M^\L\ \ \  ,\ \ \ \
\cV(v)=\phi^{9/4} \cV_4(M)
\eqn\bis
$$
we then obtain
$$
\phi_5=\phi^{3/4} \cV_4 (M)^{1/3}\ \ \ \ ,\ \ \ \cV_5(M)=\phi^{-3/4}
\cV_4(M)\ .
\eqn\quin
$$
These formulae show that $\phi_5, \cV_5(M)$ are the generalization of
${\rm Re} T$, ${\rm Re} S$ introduced in ref. \wit .

If one assumes S-T duality\duffb , the Calabi--Yau moduli are related
 to the string
coupling constants of heterotic strings and the previous theory may be used
to investigate some non-perturbative properties of heterotic string theories.
In particular, it was shown in ref.\Wdy\ that some BPS states of 5D
heterotic strings have quantum numbers related to Yang--Mills
instanton charge. On the 11D supergravity side, these states should come from
two-branes\huto\town\stromco\  wrapping around closed two-surfaces $A$
on Calabi--Yau manifolds
$$
\int_{A\times S_2} \cF
\eqn\sedi
$$
where $A$ is a two-cycle and $S_2$ is a two-sphere on $M_4$. This gives
another hint that space-time instantons and world-sheet instantons are
actually different descriptions of the same physical entities.

Let us further consider the moduli space of vector multiplets in 5D
heterotic string  theory. For a string compactified on $K3\times S_1$
this space is \gst
\foot{ Note that this manifold should not be confused with the
 $SO(1,n)/SO(n)$ manifold of n tensor multiplets coupled to $N=1$, $D=6$
supergravity\roma . In heterotic strings on $K3$ \seib ,there is only one
tensor multiplet (containing the dilaton) which yields
 the $SO(1,1)$ part. The $SO(1,n-1)/SO(n-1)$ space in $D=5$ is the Narain
 moduli
space \nara\ (modulo global modifications) of a circle compactification.}
$$
{{SO(1,n-1)}\over{SO(n-1)}}\times SO(1,1)\ ,
\eqn\dicia
$$
where the $SO(1,1)$ is the dilaton vector multiplet (including the
$B_{\mu\nu}$ field, dual to a vector) and the total number of vector
multiplets is $n$. By further reduction to $D=4$ \gst\ this manifold
becomes the special \K manifold \doubref\cremet\CDFVP\
$$
{{SO(2,1)}\over{SO(2)}}\times {{SO(2,n)}\over{SO(2)\times SO(n)}}\ ,
\eqn\dicio
$$
including $n+1$ vector multiplets. In heterotic string theory, the
$d_{\L\S\D}$ symbol is just $d_{1AB}=\eta_{AB}$, with signature
$(1,n-1)$, and vanishes otherwise. It would then seem that the dual
Calabi--Yau manifold should have the same intersection form,
possibly restricting the allowed Calabi--Yau manifolds which are dual
candidates to heterotic theories. This fact has recently been suggested
in \noi\ and verified for some dual candidates in \doubref\FHSV\kava .

 As a final remark, we would like to speculate that if $D=11$ supergravity
is taken as a serious non perturbative description of strings, then a
mechanism of supersymmetry breaking may be possible, that is the
Scherk-Schwarz mechanism\shsh\ ,
 which has been already used in heterotic string
models\fkpz . Since the gravitino gets in this case a Kaluza--Klein BPS
 mass\gmass , by string duality one would expect that this reflects in a non
perturbative gravitino mass term in the  heterotic string coupling. This
may also suggest that space-time instantons, in heterotic strings, may
induce supersymmetry breaking.

\ack{We would like to thank E. Cremmer and M. Duff for useful discussions.}

\refout
\end